\documentclass[11pt,A4]{article}
\usepackage{graphics}
\begin{document}

\begin{center}
{\large \bf INVESTIGATION OF THE  HEAT CAPACITIES OF  PROTEINS BY  STATISTICAL MECHANICAL METHODS}\\

\vspace{0.3 cm}
{\large G. Oylumluoglu$^{*1,2}$, Fevzi B\"{u}y\"{u}kk\i l\i \c{c}$^{**1}$,  Dogan Demirhan}$^{***1}$\\
$^{1}$Ege University, Faculty of Science, Department of
Physics, Izmir-TURKEY.\\
$^{2}$Mugla University, Faculty of Arts and
Sciences, Department of Physics, Mugla-TURKEY.\\

\end{center}


\begin{abstract}
In this study, the additional heat capacity which appear during
the water dissociation of the proteins that are one of the soft
materials, have been considered by the statistical mechanical
methods. For this purpose, taking the electric field E and total
dipole moment M as the thermodynamical variables and starting with
the first law of thermodynamics an equation which reveals the
thermodynamical relation between the additional heat capacity in
effective electric field $\triangle C_E$ and the additional heat
capacity at the constant total dipole moment $\triangle C_M$, has
been obtained. It is found that, the difference between the heat
capacities depends linearly on the temperature. To bring up the
hydration effect during the folding and unfolding of the proteins
the physical properties of the apolar dissociation have been used.
In the model used for this purpose; the folding and unfolding of
the proteins in the formed electric field medium have been
established on this basis. In this study with the purpose of
revealing the  additional effect to the heat capacity, the
partition functions for the proteins which have been calculated in
single protein molecule approach by A. Bakk, J.S. Hoye and A.
Hansen; Physica A, 304, (2002), 355-361 have been taken in order
to obtain the free energy. In this way, the additional free energy
has been related to the heat capacities. By calculating the heat
capacity in the effective electric field $\triangle C_E$
theoretically and taking the heat capacity at constant total
dipole moment $\triangle C_M$ from the experimental data, the
outcomes of the performed calculations have been investigated for
Myoglobin and other proteins.

\vspace{0.3 cm} {\small Keywords: Thermodynamics, Classical
statistical mechanics}


\vspace{0.3 cm} {\small $^*$ Corresponding
Authors:e-mail:oylum@sci.ege.edu.tr, Phone:+90 232-3881892
(ext.2363)
Fax:+90 232-3881892\\
\\
} \vspace{0.3 cm} {\small $^{**}$ e-mail:fevzi@sci.ege.edu.tr,
Phone:+90 232-3881892 (ext.2846)
\\
} \vspace{0.3 cm} {\small $^{***}$ e-mail:dogan@sci.ege.edu.tr,
Phone:+90 232-3881892 (ext.2381)
\\
}
\end{abstract}

\section{Introduction}
\label{intro} Now a days, a progress in the direction of
investigation of the soft materials by the methods of statistical
physics is realized. It is not possible to think a way of life
without the soft materials; all biological structures, the
molecules of the genetic code, proteins and membranes contain soft
materials. Soft materials are the fundamentals of the life and an
important component of the future technological developments.

Proteins is a common name for the complex macromolecules which are
formed by unification of a great number of amino-acids and which
play unabandonable roles in the life times of all the living
creatures. When considering the physical problem of formation of
the structure of the natural proteins, one must keep in mind that
proteins are quite different molecular objects. As a physical
object the protein molecule is rather a big polymer composed of
thousands of atoms and is such a soft material that from the
physical point of view it is a macroscopic system.

Every single atom forming the proteins occupies a place at a
certain position as in the crystals. However, unlike to the
crystals the position of each atom is unique to itself with
respect to the neighboring atoms. Consequently, proteins represent
a macroscopic system which is different from the crystals such
that here the order is not periodical. A protein could be
synthesized as a linear heterogeneous polymer which is formed by a
genetically determined order where the remainders of different
amino-acids are bounded to each other by peptide bounds. This
polypeptide chain folds to a self unique conformation which is
completely determined by the information in the ordering of the
remainders of the amino-acids. The folding of the polypeptide
chain to a three dimensional structure, principally is a
reversible process and this depends on environmental conditions.
In other words, thermodynamically, this process is passing from an
unfolded state to another rather ordered macroscopic state. This
case is named as the natural state of the protein molecule.
\section{Model}
\label{sec:1} For different aspects of the folding and unfolding
transitions of the proteins various models have been proposed
\cite{{Hensen},{Hensen2}}. Simple but yet a notable model among
then is the zipper model which defines the helix chain. In another
model, the idea of representing the solvent as dipoles in the
unfolding of proteins was given by the work of Warshel and Levitt
and later on in the applications of Rushell, Fan and Avbelj
\cite{Frank,Bakk}. In order to model the effect of the folding of
the proteins Bakk asked for the inclusion of important physical
properties in apolar dissolution. For this purpose he started by
choosing the water molecules as classical dipoles \cite{Madan}.
\subsection{First Law of the Thermodynamics for the Proteins}
\label{sec:2} By taking the electric field E and the total dipole
moment M as thermodynamical variables the first law of
thermodynamics could be written for the proteins:
\begin{equation}\label{eq:1}
    dU=TdS-MdE .
\end{equation}

Here, the electric field is not an external field but is used to
model the ice-like behavior of the water molecules around
non-polar surfaces. The electric field is a result of the
effective behavior of the non-polar solvent applied in the
unfolding of the protein. By considering the proteins as soft
materials, the thermodynamical relation between the heat capacity
at effective electric field $\triangle C_E$ and the heat capacity
at constant total dipole moment $\triangle C_M$ has been put
forward.  Re-arranging Eq.(1) one could write
\begin{equation}\label{eq:2}
    dQ=dU+MdE .
\end{equation}

To constitute such a relation, first of all the internal energy U
is written as the total differential of the variables T and E and
then substituted in Eq.(2) giving
\begin{equation}\label{eq:3}
 dQ=(\frac{\partial U}{\partial T})_E
 dT+[M+(\frac{\partial U}{\partial E})_T]dE .
\end{equation}

When the relevant heat capacities for constant M and E are written
down in Eq. (3) one obtains:
\begin{equation}\label{eq:4}
 \triangle C_{M}=\triangle C_E+[M+(\frac{\partial U}{\partial
 E})_T](\frac{\partial E}{\partial
 T})_M
\end{equation}
with the aim of expressing the term in the right hand side of
Eq.(4) in terms of experimentally measurable quantities, Eq.(4)
has been written in a different form, the total differentials of
U(T,E) and S(T,E) have been calculated, the partial
differentiations on both sides of equation have been performed and
as a result the following expression has been found
\begin{equation}\label{eq:5}
 \triangle C_{M}=\triangle C_E+T(\frac{\partial M}{\partial
 T})_E(\frac{\partial E}{\partial
 T})_M .
\end{equation}
Now, using properties of the total differential and also Maxwell
relations constituted for the proteins
\begin{equation}\label{eq:6}
 \triangle C_{M}-\triangle C_E=-T(\frac{\partial M}{\partial
 E})_T[(\frac{\partial E}{\partial
 T})_M]^2
\end{equation}
has been obtained.

Eq.(6) reveals the thermodynamical relation between the heat
capacity at effective electric field $\triangle C_E$ and the heat
capacity at constant total dipole moment $\triangle C_M$. It is
seen in the equation that; the difference of the heat capacities
depend linearly on the temperature. $\triangle C_E$ in Eq.(6)
could easily be calculated theoretically and if $\triangle C_M$ is
measured experimentally, one could get a means of investigating
the system in terms of thermodynamical quantities.
%
\subsection{Single Protein Approach}
Let us suppose that each atom has a dipole moment
\overrightarrow{p} and interatomic interactions could be
neglected. In the absence of an electric field each of the
molecules has a continuous and randomly oriented dipole moment.

On the non-denatured protein surface, in the positive bound
charges (ions) of the water molecules an ice-like construction by
the solvent effect becomes a matter of question. This in turn
leads to the orientation of the dipole moments of the polar water
molecules in the direction of field as if an external field were
applied on the water molecules. This rigidness is for providing
the water molecules to stay around the apolar surfaces. Thus this
is not a real external field but reflects the characteristics of
the solvation of the water molecules around the positive ions the
non-polar surface in the protein folding.

In this case the force on each dipole solely originates from the
produced electric field. Since each atom experiences thermal
agitation, the energy distributions of these atoms could be
investigated by Classical Maxwell-Boltzmann Statistics
\cite{Greiner}. The number of atoms $dN$ in the energy internal
$U$ and $U+dU$ is written as;
\begin{equation}\label{eq:7}
 dN=C \exp(-U \beta)dU
\end{equation}
where $\beta=1/k T$, k is the Boltzmann constant, T is the
absolute temperature and C is a constant. $U$ being the energy of
the proteins, the energy gained by the dipole due to the applied
field is given by
\begin{equation}\label{eq:8}
 U=-\overrightarrow{p} \cdot \overrightarrow{E} .
\end{equation}
Let the angle between the axis of the dipole and the electric
field be $\theta$, then change in the amount of the electrical
potential energy that the dipole possessed is obtained as
\begin{equation}\label{eq:9}
 dU=pE \sin \theta d \theta .
\end{equation}
When $dU$ is substituted in Eq.(7), the number of atoms in the
energy interval $U$ and $U+dU$ could be obtained in the following
form
\begin{equation}\label{eq:10}
 dU=C \exp(pE \beta \cos \theta)pE \sin \theta d\theta .
\end{equation}
If total number of atoms per unit volume is defined as $N=
\int^\pi_0 dN$, then one writes for the total dipole moments as
\begin{equation}\label{eq:11}
 M=N <m> = \int^\pi_0 m \cos \theta dN
\end{equation}
where $<m>=\overline{m}$ which is the average dipole moment per
atom in the direction of the formed electric field. When the
equation expressing the total number of atoms is substituted in
Eq.(11)  the          solution in terms of the Langevin function
is found as
\begin{equation}\label{eq:12}
 M= m L(a)
\end{equation}
where $pE \beta = a$.
\subsection{Hydration Effect to the Single Protein Approach}
In addition to the energy coming from the external field, in order
to introduce the pair interactions to the model pairing term must
be added. For the purpose of obtaining an effective field in
$U_{e}$ if two equations written initially are combined for a
water molecule the effective field for this case becomes
\cite{{Bakk},{Madan}}
\begin{equation}\label{eq:13}
 U_{e}=U_{DPE}+U_{MF}.
\end{equation}

Substituting the mean-field solution of the hydration effect in
the single protein approximation, one obtains for the dipole
moment \cite{{Bakk},{Bruscolini}}
\begin{equation}\label{eq:14}
 m=\int^{2 \pi}_0  d \phi \int^\pi_0  d \theta \cos \theta \sin \theta \exp (-\beta U_{e}).
\end{equation}

Supposing that, distribution of the electric dipole moments has
canonical distribution forms and making use of the fact that the
orientations of p vectors stay in the intervals $\theta +d \theta$
and $\phi + d \phi$,  from the definition of the solid angle $d
\Omega = \sin \theta d \theta d\phi$, one gets the partition
function for a protein as
\begin{equation}\label{eq:15}
 z=\int \int \exp (-\beta U_{e}) d \Omega.
\end{equation}
When solid angle $d \Omega$ and the explicit expression for
$U_{e}$ are substituted in the integral, one gets
\begin{equation}\label{eq:16}
 z(T,E,N=1)=4 \pi \exp^{-\frac{\beta b m^2}{2}} \frac{1}{\beta \epsilon_e} {\sinh \beta
 \epsilon_e} .
\end{equation}
By making use of the definition for the average dipole moment
$<m>=m/z$; it is obtained in the form $<m>= L(\beta
\epsilon_{e})$. Here the effective energy is given by $
\epsilon_{e}=\epsilon + b <m>$. The partition function of the
system is the multiplication of N number of single particle
partition functions.

In thermodynamical investigation of the protein system, in order
to introduce the chemical potential $\mu$ (here takes the place of
pH), one has to take the system in a macro-canonical ensemble. In
this study, the macro-canonical ensemble is obtained from the
canonical ensemble. Here in addition to be heat bath, it is
supposed that the system is in a particle reservoir and if this
reservoir is large, in the equilibrium, the particles reach an
average value.

In the first approximation, if the expression found in Eq.(16) for
the canonical ensemble is extended for the macro-canonical case,
without putting any limit on the particle number \cite{Greiner}
one could use the relation
\begin{equation}\label{eq:17}
 Z(T, E, \mu)=\sum^\infty_{N=0}\frac{1}{N!}[\exp(\frac{\mu}{k T})z(T, E,
 N=1)]^N .
\end{equation}

The partition function of the undistinguishable particles  which
is given by Eq.(17) is obtained as
\begin{equation}\label{eq:18}
Z(T, E, \mu)=\exp[\exp(\frac{\mu}{k T})z(T, E,
 N=1)]
\end{equation}
where the expression $\zeta=\exp(\beta\mu)$ corresponds to the pH
of the system and is the weight factor. In this case,
\begin{equation}\label{eq:19}
<U>=-\frac{1}{Z}\frac{\partial}{\partial\beta}(Z)_\zeta,_E
\end{equation}
is written for the energy of the system. Here, the expression $
\Phi(T,E,\mu)=-\frac{\ln Z(T,E,\mu)}{\beta}$  is the
thermodynamical potential. After the partition function obtained
by the substitution of Eq.(16) in Eq.(18) the thermodynamical
quantities of the system are found as follows:

Entropy is given by the formula
\begin{equation}\label{eq:20} \nonumber
S( T, E, \mu)=-(\frac{\partial\Phi}{\partial T})_E,_\mu
\end{equation}
which then leads to
\begin{eqnarray*}\label{eq:20}
S(T, E,\mu)=k[\ln[\exp(\frac{4
\exp\beta(\mu-\frac{bm^2}{2})}{\epsilon_e \beta})] \\
+\frac{\exp \beta(\mu-\frac{b m^2}{2}) [-4 \pi \epsilon_e \beta
\cosh(\epsilon_e \beta)} {\beta \epsilon_e}+ \\ \frac{2 \pi
(2+\beta b m^2) \sinh(\epsilon_e \beta)]} {\beta \epsilon_e}]
\end{eqnarray*}
total dipole moment is given by the formula

\begin{equation}\label{21}
   M(T, E, \mu)=-(\frac{\partial\Phi}{\partial E})_T,_\mu
\end{equation}
which then leads to

\begin{eqnarray*}\label{eq:21}
M(T, E, \mu)=\frac{4 \pi \exp(\beta(\mu- \frac{b m^2}{2}))} {\epsilon_e^2 \beta^2} \\
(\epsilon_e \beta \cosh(\epsilon_e \beta)-\sinh(\epsilon_e \beta))
\end{eqnarray*}
and lastly the particle number is given by the formula
\begin{equation}\label{eq:22}
N(T, E, \mu)=-(\frac{\partial\Phi}{\partial \mu})_T,_E
\end{equation}
leading to
\begin{eqnarray}
\nonumber
 N(T,E,\mu)=\frac{4 \exp(\beta(\mu-\frac{b m^2}{2})) \pi
\sinh(\epsilon_e \beta)}{\epsilon_e \beta} .
\end{eqnarray}

\section{Variation of the Additional Heat Capacity with Temperature}
\label{sec} In this study the variation of $\triangle C_M$ with
temperature has been investigated. In the macro-canonical ensemble
the additional heat capacity in the effective electric field for
the protein system is calculated from the expression
\begin{equation}\label{eq:23}
\triangle C_{E}=(\frac{\partial U}{\partial T})_\zeta,_E -
\frac{1}{T}(\frac{\partial N}{\partial \mu})_E,_T [(\frac{\partial
U}{\partial N})_T,_E]^2 .
\end{equation}
On the other hand, the variation of the internal energy with
temperature is given by
\begin{equation}\label{eq:24}
(\frac{\partial U}{\partial T})_\zeta,_E=(\frac{\partial
U}{\partial T})_E,_N+(\frac{\partial U}{\partial N})_T,_E
(\frac{\partial N}{\partial T})_\zeta,_E .
\end{equation}
For $(\frac{dU}{dN})_{T,E}$ in Eq.(23)
\begin{equation}\label{eq:25}
(\frac{\partial U}{\partial N})_T,_E=\frac{(\frac{\partial
U}{\partial T})_\zeta,_E-(\frac{\partial U}{\partial
T})_E,_N}{(\frac{\partial N}{\partial T})_\zeta,_E}
\end{equation}
is substituted which is attained from Eq.(24).

$\triangle C_E$ is substituted in Eq.(6). $M, S $ and $N $ that
are necessary in Eq.(6) and Eq.(25) have been taken from Eq.(21),
Eq.(20) and Eq.(22) respectively. The explicit forms of the above
expressions are lengthy and thus are not written down.

In order to compare $\triangle C_M$ which is found theoretically
with the experimental values of $\triangle C_M$ they are shown on
the same plot in Fig(1). The dashed curves represent the
experimental study and the solid curves represent the theoretical
result. In this semi-phenomenological theory, the physical
quantities $\epsilon$ , $ b $ and $\mu$ curves have been
determined by fitting to the results of Privalov and Bakk .

\begin{figure}[t]
\resizebox{0.45\textwidth}{!}{%
  \includegraphics{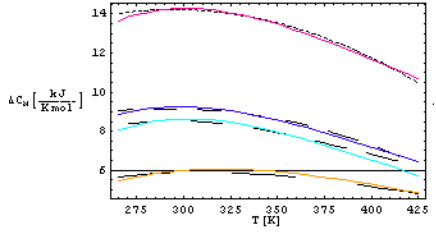}
}
\caption{Graph of the variation of the additional heat capacities
with temperature.}
\label{fig:1}       
\end{figure}

Red (Myoglobin), blue (Lysozyme),  green (Cytochrome) and Yellow
(Ribonuclease) solid curves represent the variation of the
additional heat capacity difference with temperature obtained from
theoretical result.
\begin{table}[h]
\caption{ Additional heat capacity values corresponding to
transition temperature for four different proteins.}
\label{tab:1}       
\begin{tabular}{llll}
\hline\noalign{\smallskip}
Color & Protein & AHCD$(\frac{kJ}{Kmol})$ & Trans. Temp.$(K)$ \\
\noalign{\smallskip}\hline\noalign{\smallskip}
Red & Myoglobin & 14.3 & 301 \\
Blue & Lysozyme & 9.3 & 298 \\
Green & Cytochrome & 8.6 & 302 \\
Yellow & Ribonuclease & 6.1 & 320 \\
\noalign{\smallskip}\hline \\
\end{tabular}
\begin{tabular}{l}
~~A   H   C   D   : ~Additional Heat Capacity Differences \\
\noalign{\smallskip}\hline
\end{tabular}
\vspace*{0.1 cm}  
\end{table}

\section{Conclusions}
\label{sec} In this study a theory of the heat capacity depending
on electrostatic field and constant dipole moment in the
transition between the fold and unfold states of the protein
molecule according to Bakk \cite{Bakk} model, has been presented.
The foresight of the performed calculations have been discussed by
taking into account the experimental studies of Privalov and
Makhatadze on Myoglobin, Lysozyme, Cytochrome and Ribonuclease in
the interval $275-350K$ \cite{Privalov,Privalov2}.

Biological and other macromolecular systems exhibit great
variations in heat capacity with respect to temperature. The
unfolding of the globular proteins in water is a classical
example. Change in the environmental conditions such as
temperature, increase or decrease in pH, interaction of the
solution dissolvers with protein molecule groups, gives rise to
the change in the structure of the proteins.

When the conditions return to their initial state these molecules
return to their former structures. The folding and then unfolding
of these molecules occur by a thermodynamical mechanism. When the
graph of the variation of the additional heat capacities with
temperature is examined one could state that the behavior of the
proteins exhibits approximately an universal behavior. The apex in
the heat capacity curve indicates the denaturation transition. For
each of the four proteins; additional heat capacity difference and
transition temperatures $T_{t}$ correspond to different values and
the numerical values have been presented in Table 1. For every
protein in native state the heat capacity is less than the value
in the denature field. The transition between the native and
denature state is a first order transition.

An additional reason of the observed increase in the heat capacity
is the hydration effect. The transformation of the apolar solvents
into water, takes place with a considerable increase in the heat
capacity. A conclusion drawn here is that; depending on the
unfolding of the proteins, the additional increase in the heat
capacity is a consequence of the contact of the internal groups
during the unfolding of the proteins. The hydration effect
originates from the low solubility of the apolar compounds in
water.

One of the characteristics of the folding and unfolding of the
proteins is that, the temperature interval of the process depends
on environmental conditions and particularly on the pH values of
the solution. This task is supplied by the term $\zeta=\exp( \beta
\mu)$ in the partition function. Variations in the enthalpy
difference relevant to pH will be presented in a forthcoming
article.

\section{Acknowledgment}
\label{sec} One of the authors, G.O. thanks to Prof. Dr. Sener
Oktik, rector of the Mugla University for his contributions as
well as his support to stay in Ege University, Physics Department
where this study has been carried out.

%

%
%

\end{document}